# Artificial Fun: Mapping Minds to the Space of Fun

**Soenke Ziesche**
Singularity University
sziesche@gmx.net

**Roman V. Yampolskiy**
University of Louisville
roman.yampolskiy@louisville.edu

**Abstract**
Yampolskiy and others have shown that the space of possible minds is vast, actually infinite (Yampolskiy, 2015). A question of interest is "Which activities can minds perform during their lifetime?" This question is very broad, thus in this article restricted to "Which non-boring activities can minds perform?" The space of potential non-boring activities has been called by Yudkowsky "fun space" (Yudkowsky, 2009). This paper aims to discuss the relation between various types of minds and the part of the fun space, which is accessible for them.

**1 Introduction**
We motivate the relevance of the question "Which non-boring activities can minds perform?" as follows: There is optimism for a significant extension of the lifetime of humans in the near and medium future. Yet, some people expressed concern that there is not enough accessible fun space, thus much longer lives would not be fulfilling and mostly boring (Dvorsky, 2012).

In addition, future scenarios can be envisaged when humans are enhanced by intelligence amplification and other transhuman features, potentially supported by friendly AI. Further scenarios include creating substrate-autonomous persons, e.g. through uploading or whole brain emulation (Bostrom, 2014). In the latter case lifetime could increase almost indefinitely, and also former unknown sensations could be perceived. Also for this option, concerns have been raised that it will turn out to be boring for such enhanced humans due to the lack of remaining intellectual challenges. In contrast, others believe the fun space is large enough for either scenario (e.g. Moravec, 1988; Yudkowsky, 2009; Bostrom, 2008)

Yet another view by Pearce is that in a parallel development boredom will be "neurochemically impossible" through advanced technology (Pearce, 2012; Yampolskiy, 2015). Therefore, although human minds are only a small subset of the space of possible minds, for this type of minds the question about sufficient non-boring activities, i.e. enough accessible fun space, is very important.

In addition, we are addressing the question of accessible fun space also for non-human minds by aiming to contribute to one aspect of the field of "intellectology" (Yampolskiy, 2015). Yampolskiy has introduced "intellectology" as a new area of study in order to turn away from a human-centric view regarding minds and to examine in more detail features of any possible minds. This universe of possible mind designs is actually vast and certainly merits investigation; especially given the possible scenario that one group of minds, that is minds of AI agents, may cohabit with us in the not so far future.

This paper is structured as follows: In the following section we provide an introduction to the space of minds. This is followed by an introduction to the space of fun. In the central section we analyze relationships between the space of minds and the space of fun.

**2 The space of minds**
When we reason about minds we tend to think of human minds only. This is because of the anthropomorphic bias. However, in addition, there are other minds, which we encounter on earth, the minds of higher order animals, and then there are various minds, which we can imagine as possibility and perhaps even more beyond our imagination ("unknown unknowns"). It has been shown that the space of possible minds is vast (e.g. Sloman, 1984; Goertzel, 2006; Yudkowsky, 2008; Yampolskiy, 2015).

Examples for potential minds could be human-designed AI minds, self-improving minds, a combination of minds constituting itself a mind and many more. There have been several attempts to classify the space of minds (Yampolskiy, 2015). In fact, the space of human minds forms only a tiny subset within the universe of possible minds (Yudkowsky, 2008). The space of possible minds can be considered as the set of possible cognitive algorithms. Based on this and on the limited number of cognitive algorithms, which human minds can potentially perform, it can be concluded that the majority of possible minds is more intelligent than human minds. Yampolskiy and Fox describe this insight as another example of a Copernican Revolution, i.e. a revision of the view that humanity is central, which in this case refers to minds (Yampolskiy & Fox, 2012).

Yampolskiy shows that the set of mind designs is infinite and countable and that all minds can be generated sequentially by a deterministic algorithm, based on a variant of the Levin Search (Yampolskiy, 2015). By linking a mind to a number, Yampolskiy also demonstrates that "[…] a mind could never be completely destroyed, making minds theoretically immortal. A particular mind may not be embodied at a given time, but the idea of it is always present." (Yampolskiy, 2015). The fact that minds are countable will be useful when establishing relations with the fun space.

**3 The space of fun**
Regarding the goals and the motivation of minds Bostrom distinguishes between two theses, the orthogonality thesis and the instrumental convergence thesis (Bostrom, 2014). The former states that "more or less any level of intelligence could in principle be combined with more or less any final goal.", while the latter proposes that there is a rather limited number of instrumental values which many intelligent agents share.

The focus of this paper are those goals of minds that are related to the discovery of novelties and to having fun, which is the opposite of boredom. Yudkowsky proposes a link between novelty and fun: "Novelty appears to be one of the major keys to fun, and for there to exist an infinite amount of fun there must be an infinite amount of novelty." (Yudkowsky, 2009). We shall specify this further below.

Omohundro defines four drives for AI minds: Efficiency, self-preservation, acquisition, and creativity (Omohundro, 2007, Omohundro, 2008). Three of these drives address novelty: The creativity drive is about novelty by definition (leads to the development of new concepts, algorithms, theorems, devices, and processes), but so are the efficiency drive (aiming for novel algorithms and methodologies) and the acquisition drive (aiming for novel resources). Only the self-preservation drive is linked more indirectly to novelty, which has to be relative to a particular self. Hence, without self-preservation everything might be seen as novel to some new version of that self. However, Omohundro does not mention "fun" to be linked to any of these AI drives.

We know that at least human minds aim for accessing fun space. Therefore, the subset of all minds, with the distinction to have the objective of having fun, is at least the set of all human minds, but potentially it is bigger. Yet, when pondering over such questions it is always critical to beware of the anthropomorphic bias:

- It could well be possible that fun is specific to humans only and other minds do not aim for fun. The actions of these minds are still likely to be steered by the above drives, yet they may not aim for fun as an epiphenomenon and may be satisfied by pursuing boring activities as long as these serve their goals (Yet, we seem to observe at least some higher order animals having fun.)

- If there are other minds that aim for fun, this "fun" may look extremely different from what is fun for humans and may well be unimaginable for humans. Both, the sensation that other minds "feel" as fun as well as the activities that trigger this sensation could be very different, e.g. activities that are considered tedious by humans may cause for other minds fun-type sensations as epiphenomenon.

- There may be certain minds, who do aim for fun, but apply "wireheading" methods, which is obtaining the reward through direct stimulation without otherwise required productive behavior. Hence, drugs and/or technologies provide them effortlessly and permanently with what they consider as fun (Yampolskiy, 2014).

**What is the relation between novelty and fun?**
It seems that fun is useful to motivate exploration, innovation and discovery. Stanley and Lehman show that novelty search is more successful than objective-based search (Lehman & Stanley, 2015). Therefore, striving for fun of human minds could be explained by evolution. Novelties are critical for human development, and fun is, at least sometimes, an epiphenomenon while pursuing novelties or at the moment when discovering novelties. However, it can be easily shown that, at least for human minds, discovering novelties and having fun is not the same, since there are numerous novelties that are boring for humans; or a specific novelty could be fun for one human mind, but boring for others.

Schmidhuber proposes the following options to access fun space, which we will take into account below: "There are at least two ways of having fun: execute a learning algorithm that improves the compression of the already known data […], or execute actions that generate more data, then learn to compress/understand the new data better." (Schmidhuber, 2010)

It is speculation whether this also applies to non-human minds. For example, does the mind of a "paperclip maximizer" (Bostrom, 2003) accesses fun space when discovering novelties through the creativity drive or the efficiency drive (e.g. finding time- or energy saving ways to produce paperclips) or the acquisition drive (e.g. finding new resources to be converted into paperclips)?

**What is the size of the fun space?**
Since potential knowledge is infinite also the amount of novelties that could potentially be gained by a mind is infinite, yet limited by time and other resources.

But as we stated, not every processing of novelties triggers fun as an epiphenomenon. Given the infinity of the knowledge space it may be assumed that the potential fun space is vast. In this regard Yudkowsky posed critical questions: "How much fun is there in the universe? Will we ever run out of fun? Could we be having more fun?" (Yudkowsky, 2009). We pose an additional question here, which brings us to the central part of this paper, i.e. we attempt to establish relations between the space of minds and the space of fun: Which minds have (in principle) access to which part of the fun space?

**4 Relations between the space of minds and the space of fun**
For the formalization we assume that both, all potential minds (Yampolskiy, 2015) can be sequentially generated as well as all possible information strings (Yampolskiy, 2013; Schmidhuber, 2010; Lehman & Stanley, 2015). We also use Yampolskiy's insight that knowledge is discovered by finding and applying an efficient algorithm to an inefficiently stored information string (Yampolskiy, 2013).

In order to investigate the fun space for certain minds we define a Boolean function ("fun") with two parameters, a string representing a mind and another string representing information/knowledge as follows:

**fun(mind string, info string) = {true, false}**

The verbal description of this function would be that it becomes true if for a certain mind, represented by a numeric string, the processing of certain information, also represented by a numeric string, triggers fun as an epiphenomenon. For all other combinations of mind strings and information strings the function is false. The formalization is as follows:

fun(x, y) = true, if

[The mind x has the ability/intelligence to turn the info string y efficiently into knowledge (**access**).]

AND

[[The mind x had never turned that specific info string y into knowledge before. (**novelty fun**)]

OR

[The mind x considers the computation process of info string y into knowledge as fun, regardless whether it has accessed the info string y before or not. (**process fun**)]]

This means, the **access** to the knowledge represented by the information string is the precondition for fun. Although the set of human minds is very homogeneous, the extent of knowledge, which humans access during their lifetime varies significantly because certain minds can not access 1) certain knowledge for intellectual reasons and or 2), even if they are intelligent enough, there could be resource or time restrictions:

- <u>Available time and resources</u>: Some minds have to devote most of their resources and time for "survival", i.e. have less resources and time for fun. An example would be a rich person in a developed country vs. a poor person in a developing country.

- <u>Life span</u>: Minds are embodied for different durations, i.e. have different life spans, which imposes time restrictions on accessing knowledge.

- <u>Time of embodiment</u>: The accessible knowledge was for a particular mind in Stone Age quite different from one, who is embodied in 21$^{st}$ century.

We have shown that accessible knowledge varies immensely between minds and depends on intelligence, time, resources etc to efficiently turn certain information strings into knowledge. Therefore, certain knowledge can neither be part of the process nor the novelty fun space of these minds due to their inability to access it.

After it has been established that a mind x can access the knowledge represented by an information string y, at least one of two other assertions has to be true, so that fun is involved: New knowledge is ensured 1) by retrieving a new info string y (**novelty fun**) or 2) merely due to the computation process that is applied to info string y (**process fun**).

These two facets of fun are motivated by Schmidhuber's "two ways of having fun" (Schmidhuber, 2010) and by Yampolskiy's insights about the relation between information strings and knowledge (Yampolskiy, 2013): "Since multiple, in fact infinite, number of semantic pointers could refer to the same string that means that a single string could contain an infinite amount of knowledge if taken in the proper semantic context, generating multiple levels of meaning." Examples in Table 1 illustrate the different cases. The fun space is marked with grey background:

Table 1: Types of fun

| Mind x accesses info string y | New info string y | Previously accessed info string y |
|---|---|---|
| **New computation process** | **Novelty and process fun**<br>- To watch for the first time a 3D movie.<br>- To proof that there are infinite prime numbers (Euclid).<br>- To perform qualia surfing, i.e. to experience new qualia (as an uploaded mind, see Loosemore). | **Process fun**<br>- To find a different proof that there are infinite prime numbers (Euler)[1].<br>- To listen to a known song again in different context or mood.<br>- To look at a painting again in different context or mood. |
| **Trivial/routine computation process** | **Novelty fun**<br>- To discover a new species of beetle in the rain forest.<br>- To watch a new episode of a cherished TV series.<br>- To do sightseeing in a new city. | - To clean the dishes. |

To elaborate on some examples:

- **Novelty fun:** To discover a new species of beetle in the rain forest.

---

[1] Euler lived around 2,000 year later than Euclid and knew about Euclid's proof, i.e. the knowledge about the infinity of prime numbers had been accessed before, thus no involvement of "novelty fun" for Euler, but his proof was different, thus potential involvement of "process fun" for him.

In this table "trivial/routine process" means that no innovative computation process was required to gain the new knowledge. A biologist may have conducted several expeditions to the Amazon rainforest and at a certain day by using her typical (therefore, trivial) methodology and basically her eyes, she discovers a new species of beetle, which, on a very basic level, is accessing a new information string (which may have never been accessed by a human mind before) and for her this is equivalent with having **novelty fun**.

- **Process fun:** To look at a painting again in different context or mood.

It is not uncommon to hear people enjoying looking at the same painting again and again, e.g. revisiting a particular museum regularly for this purpose. Except for the case that they discover new details in the painting, they, on a very basic level, access the same information string and by doing so, even repetitively, they have **process fun**. The explanation is provided by Yampolskiy: The same information string can contain different knowledge and multiple levels of meaning (Yampolskiy, 2013).

- **Novelty and process fun:** To perform qualia surfing as an uploaded mind.

Loosemore defines "qualia surfing" in the context of uploading as "collecting new experiences by transferring our consciousness back and forth between different substrates" (Loosemore, 2014). One of his several examples is to change "into a body that can swim in the atmosphere of Jupiter […], while you stargaze through telescopic eyes that can see beyond the visible spectrum". He explains that even if qualia surfing is restricted to the real world there is vast space of new experiences, which would be even larger if virtual worlds are added. For a person experiencing certain qualia for the first time **novelty as well as process fun** may be involved.

A less futuristic example is to watch for the first time a 3D movie. The computation process, via 3D glasses, is new as well the information string, i.e. content of the movie; given the person has not watched it in 2D before.

It has to be highlighted that it is critical that the function fun(mind string, info string) has **two parameters**. Fun does not only depend on the accessed information string, but very much on the individual mind. For example, there are obviously human minds who neither enjoy watching movies nor discovering new species of beetles etc.

A weakness of human minds and therefore a special case is **forgotten knowledge**. In contrast, it is very likely that more sophisticated minds will have a nearly perfect memory. When human minds regain forgotten knowledge both could be possible, to have fun or not: It could be fun to watch a movie again after a long time because one had forgotten the storyline. It may not be fun, but frustrating to spend time re-learning vocabulary of a language, which one had not practiced for some time. In both cases the same information string is accessed by the same computation process when the knowledge had been computed for the first time.

Based on Yampolskiy's approach and other mathematical truths it is evident that there are **infinite information strings**, which can be turned through **infinite computation processes** into **infinite knowledge** (Yampolskiy, 2013). This means every access to new knowledge and every new computation of existing or new knowledge may or may not involve novelty or process fun respectively. Therefore, the questions remain:

**Is the fun space also infinite? If yes, are human minds able to access an infinite subset of this fun space?**

One constraint was mentioned above: Some knowledge remains hidden from human minds due to insufficient intelligence, time and resources. Those parts of fun space that are linked to the novelty and processing of this inaccessible knowledge are thus also inaccessible for human minds. We look at an excerpt of a random sample truth table for the function fun(mind string, info string), which is infinite and countable since both parameter sets are infinite and countable.

Table 2: Random sample truth table for the function fun

| Sample Truth table fun(x,y) | Human mind (t/f) | Info 1 | Info 2 | Info 3 | Info 4 | Info 5 | Info 6 | Info 7 | Info 8 | Info 9 | … |
|---|---|---|---|---|---|---|---|---|---|---|---|
| Mind 1 | f | f | t | f | t | f | t | t | f | f | |
| Mind 2 | t | f | t | t | f | f | f | t | f | t | |
| Mind 3 | f | f | f | f | t | f | f | t | t | f | |
| … | | | | | | | | | | | |

The overall fun space would be infinite if the truth table of this function would contain an infinite number of the value "true". This we cannot prove. It is possible that the number of the value "true" is finite, which implies that in this case the number of the value "false" must be infinite.

For the second question above, we look at specific minds, human minds in particular, and their accessible fun space. Therefore, we added the second column in the above truth table, which filters human minds out of the space of all minds. In this example "Mind 2" happens to be human, and this mind would have infinite fun space if that row contained an infinite number of the value "true". The above question whether human minds have access to an infinite subset of the fun space would be affirmative, if the following statement was true:

$$\forall x: human\_mind(x) = t \land |\{y \mid fun(x, y) = true\}| = \infty$$

We do not have a formal answer to this. In fact it is almost certain that not all human minds have access to infinite fun space. However, we can observe that the accessible fun space has risen in history and is currently enormous, which is linked to what Kurzweil describes as the "Law of Accelerating Returns" (Kurzweil, 2005). According to this law the rate of change in certain developing systems, for example technologies, has shown to increase exponentially. As already mentioned earlier the access to the fun space of humans who lived hundreds or thousands of years ago was much more restricted than it is for present humans. Only thanks to technologies many novelties can be experienced. And also thanks to technologies we have more time at hand, which does not have to be spent on life-sustaining activities. Therefore, we may assume that future generations can access even larger parts of the fun space.

**Fun for human minds with extended life span**
The concern has been expressed that humans with an extended life span will run out of fun (Dvorsky, 2012). Yet, an extended life span enables that one limitation above is alleviated even further (time). Therefore, all else equal (e.g. the intelligence of this ageing mind stays the same) this human mind has the potential to gain additional knowledge during the additional time of embodiment, which may entail novelty and/or process fun.

It is hard to speculate how the access to fun space will be for humans who live for thousands or millions of years, but so far life expectancy has been increased successively by smaller time frames only. It is likely that more radical extensions would be accompanied by other developments to the human phenotype, which we discuss below.

**Fun for human minds enhanced by intelligence amplification and other transhuman features, potentially supported by friendly AI**
Similarly, concern has been raised that humans enhanced by transhuman features and/or supported by friendly AI may run out of fun: "If we self-improve human minds to extreme levels of intelligence, all challenges known today may bore us. Likewise, if superhumanly intelligent machines take care of our every need, it is apparent that no challenges nor fun will remain." (LessWrong Wiki, 2016).

This argument addresses mainly the above defined "process fun", i.e. previous challenges are becoming too simple. Yet, we have stated not only the infinity of computation processes as well as the infinity of knowledge, but also shown that trivial/routine computation processes could tap fun space as long as a new information string is accessed (of which also an infinite number exists).

Therefore, even if such minds run out of "process fun", since previously challenging computation processes became much simpler and/or can be performed much faster, there should be still sufficient "novelty fun" available. Chances for "novelty fun" are actually much higher than for unenhanced human minds, since enhanced human minds can reach previously inaccessible knowledge due to 1) improved intelligence, 2) much longer life span and 3) more and other resources. Enhanced humans will likely have access to additional sensors, which provide for "novelty fun" through experiencing new qualia (Loosemore, 2014). This is in line with e.g. Bostrom and Yudkowsky, who also believe that enhanced minds will not run out of fun, given that they are embodied in a sufficiently advanced context (Bostrom, 2008; Yudkowsky, 2009). Bostrom is even optimistic regarding "process fun": "If some challenges become too easy for posthumans, they could take on more difficult challenges."

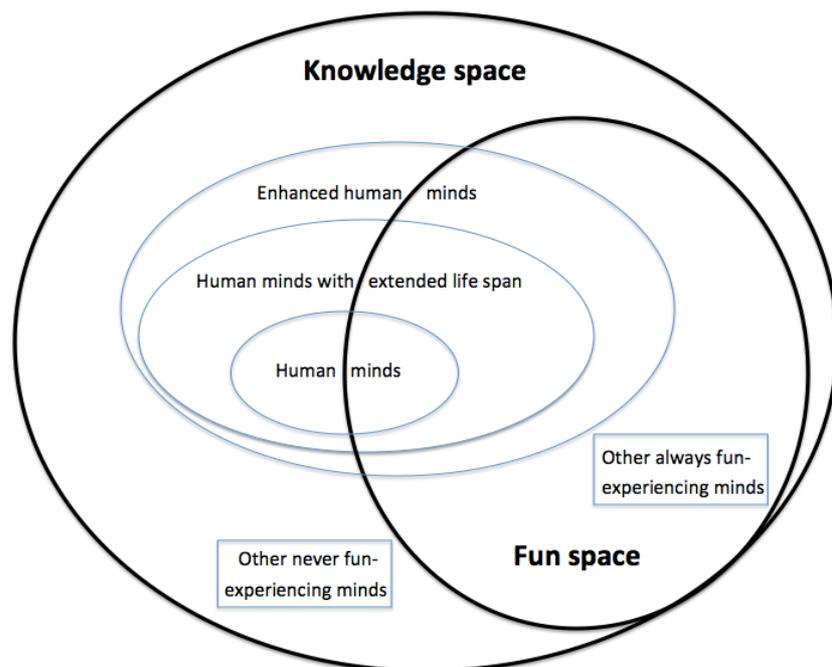

Figure 1: Access of a selection of minds to knowledge and fun space

As indicated above, another scenario in the future, which would make the above considerations obsolete, is the manipulation of human and other minds through technology, known as **wireheading**, in a way that boredom will be "neurochemically impossible" as envisaged by Pearce (Pearce, 2012). For such a manipulated mind even uninterrupted dishwashing would be continuous fun.

**Fun for AIs and other minds**
We lack evidence to what extent non-human minds can access the fun space, i.e. have a sensation of fun as epiphenomenon of any of their activities. It could be anywhere between the two extremes, which are to never experience fun (x1) or to have fun no matter which information string is accessed (x2).

$\forall$ y: fun(x1, y) = false
$\forall$ y: fun(x2, y) = true

At least for those non-human minds that are to some extent steered by the drives proposed by Omohundro (Efficiency, self-preservation, acquisition, and creativity) (Omohundro, 2007, Omohundro, 2008) we may assume that they aim for novelties. On the one hand, it would be anthropomorphism to assume that their pursuit of novelties is accompanied or motivated by any form of fun. On the other hand, if we take our definition of the mind space as the set of all possible cognitive algorithms, i.e. an infinite set, there must be other minds, apart from humans, which are able to access fun space. And considering the space of all possible cognitive algorithms, there must be even minds experiencing fun by performing activities, which have nothing to do with detecting novelties.

**5 Conclusion and future work**
Based on the groundwork of having described the possible space of minds as well as the space of fun we established a relation between the two spaces. We have defined a Boolean function fun(mind string, info string) with two parameters, one numeric string representing a mind and another numeric string representing information/knowledge.

We have shown that human minds could experience two types of fun by transforming information to knowledge: novelty fun and process fun. By means of this function we were able to formalize the critical question whether human minds are able to access an infinite subset of the fun space. It is beyond our means to give a definite answer to the question, but we provided a substantiated assumption that a large number of human minds in present time and especially in the future have access to a vast space of fun.

These findings have ethical relevance since human minds strive for fun (not merely knowledge). In history as well as still in present time there have been and there are numerous humans living a life predominantly characterized by life-sustaining activities, which do not involve much fun. Yet, ideally the human condition improves over time owing, among other things, to advances in technology. Therefore, in order to determine whether the extension of the life span and/or enhancements for humans really mean progress to the human condition it was critical to show that there will be no shortage of accessible fun space for these minds. If it had turned out that there is not sufficient fun for either human minds with extended life span or enhanced human minds such developments would probably not be desirable.

Moreover, these findings can be seen as a contribution to the new field of "intellectology" (Yampolskiy, 2015). This field is still in its early stages and provides for large areas of further research. To analyze relations between various kinds of minds and the fun space extends the traditional human-centric space of research and this is what "intellectology" is about. Further research is required on the question how to narrow down or to predict more precisely which new computation processes or which new knowledge mean fun for a certain mind, especially humans, and which are not part of the fun space despite being novel?